\documentclass[useAMS,usenatbib]{mn2e}
\usepackage{graphicx,amssym}
\newcommand{\bc}{\begin{center}}
\newcommand{\ec}{\end{center}}

\title[The formation history of
      elliptical galaxies]
      {The formation history of elliptical galaxies} 
\author[G.~De Lucia et al.]
       {Gabriella De Lucia$^1$\thanks{Email: gdelucia@mpa-garching.mpg.de},
         Volker Springel$^1$, Simon D.~M.~White$^1$, Darren
         Croton$^1$\thanks{Present address: Department of Astronomy, University
           of California, Berkeley, CA 94720. } 
        \newauthor and Guinevere Kauffmann$^1$
        \\      
        $^1$Max--Planck--Institut f\"ur Astrophysik, 
        Karl--Schwarzschild--Str. 1, D-85748 Garching, Germany}
\begin{document}


\pagerange{\pageref{firstpage}--\pageref{lastpage}} 
\pubyear{2005}

\maketitle

\label{firstpage}

\begin{abstract}
  We take advantage of the largest high-resolution simulation of cosmic
  structure growth ever carried out -- the {\it Millennium Simulation} of the
  concordance $\Lambda$CDM cosmogony -- to study how the star formation
  histories, ages and metallicities of elliptical galaxies depend on
  environment and on stellar mass. We concentrate on a galaxy formation model
  which is tuned to fit the joint luminosity/colour/morphology distribution of
  low redshift galaxies. Massive ellipticals in this model have higher metal
  abundances, older luminosity--weighted ages, shorter star formation
  timescales, but lower assembly redshifts than less massive systems.  Within
  clusters the typical masses, ages and metal abundances of ellipticals are
  predicted to decrease, on average, with increasing distance from the cluster
  centre. We also quantify the effective number of progenitors of ellipticals
  as a function of present stellar mass, finding typical numbers below $2$ for
  $M_*<10^{11}\,{\rm M}_\odot$, rising to $\sim 5$ for the most massive
  systems. These findings are consistent with recent observational results that
  suggest ``down-sizing'' or ``anti-hierarchical'' behaviour for the star
  formation history of the elliptical galaxy population, despite the fact that
  our model includes all the standard elements of hierarchical galaxy formation
  and is implemented on the standard, $\Lambda$CDM cosmogony.
\end{abstract}


\section{Introduction}
\label{sec:intro}

Elliptical galaxies are the most massive stellar systems in the local Universe
and appear to define a homogeneous class of objects with uniformly old and red
populations, negligible amounts of gas, and very little star formation.  Their
deceptively simple appearance inspired a `classical' formation scenario in
which they form in a single intense burst of star formation at high redshifts
($z\gtrsim5$), followed by passive evolution of their stellar populations to
the present day \citep{PP67,larson75}.  This so-called \emph{monolithic}
scenario successfully explains the tightness of the fundamental scaling
relations that elliptical galaxies obey, like the colour--magnitude relation
and the fundamental plane, as well as the evolution of these relations as a
function of redshift \citep{kodama98,vandokkum03}.

This classical view has been tenaciously resistant to challenges by other
theoretical models, despite numerous indications for a more complex formation
scenario.  \citet{toomre72} suggested that elliptical galaxies can form from
major mergers of massive disk galaxies.  Detailed numerical simulations
\citep{FS82,NW83} later showed that the merger of two spiral galaxies of
comparable mass can indeed produce a remnant with structural and photometric
properties resembling those of elliptical galaxies.  In more recent years, a
large body of observational evidence has been collected that demonstrates that
interactions and mergers indeed represent a common phenomenon at high
redshifts, and that these processes affect the population of elliptical
galaxies in the local Universe.  \citet{schweizer92} found evidence for bluer
colours of elliptical galaxies with increasing morphological disturbance in a
study based on a small sample with a strong bias towards isolated systems
\citep[see][]{michard04}.  Later studies using absorption--line indices have
demonstrated that a significant fraction of cluster early--type galaxies has
undergone recent episodes of star formation \citep{barger96}.  Signs of recent
star formation activity have also been detected in a number of high redshift
early--type galaxies using colours \citep*{menanteau01,vandeven03} and
absorption and emission line diagnostics \citep{treu02,willis02}.  These
results favour, at least for a part of the elliptical galaxy population, a
\emph{hierarchical} formation scenario in which larger spheroidals are
assembled relatively late from the merger of late--type galaxies of comparable
mass.  Such a bottom-up formation scenario is naturally expected for the
structure formation process in cosmologies dominated by cold dark matter.

However, despite an enormous amount of work both on the theoretical and on the
observational side, the debate concerning the two competing theories for the
formation of elliptical galaxies has remained open.  As described above, a
relatively large fraction of early--type systems shows clear evidence of
interactions, mergers, and recent star formation.  However, the data also seem
to indicate that only a small fraction of the mass is involved in such
episodes.  The latter observational result has often been interpreted as strong
evidence against the more extended star formation history naively predicted
from hierarchical models.  A related issue concerns the $\alpha$-element
enhancements observed in ellipticals.  The so--called $\alpha$-elements are
released mainly by supernovae type-II, while the main contribution to the
Fe-peak elements comes from supernovae type-Ia.  For these reasons, the
[$\alpha$/Fe] ratio is believed to encode important information on the
time--scale of star formation.  It is now a well established result that
massive ellipticals have super-solar [$\alpha$/Fe] ratios, suggesting that they
formed on relatively short time--scales and/or have an initial mass function
that is skewed towards massive stars.  The inability of early models of the
hierarchical merger paradigm to reproduce this observed trend has been pointed
out as a serious problem for these models \citep{thomas99}.

Another contentious issue is related to the significantly dissimilar evolution
of early--type galaxies in different environments predicted by early
semi--analytic models of galaxy formation \citep*{kauffmann96a,baugh96}.  Such
differential evolution is a natural outcome of the hierarchical scenario,
because present day clusters of galaxies form from the highest peaks in the
primordial density fields, leading to an earlier onset of the collapse of the
dark matter haloes and to more rapid mergers \citep{kauffmann95}.

In recent years, considerable observational progress has been made in the study
of the stellar populations of elliptical galaxies in different environments and
at different redshifts \citep[e.g.][]{thomas04,denicolo04,treu05,vandervel05}.
One has to bear in mind however that the derived ages and metallicities depend
quite strongly on the models employed in the analysis, primarily because of an
essentially unavoidable intrinsic age-metallicity degeneracy, which appears to
be stronger for older and more metal rich systems \citep{denicolo04}.  The
situation is especially unclear for `field' galaxies, since here the very
definition of `field' is fraught with ambiguities.  Significant systematic
uncertainties are also present in theoretical models used for studying
galaxy formation, where poorly understood physical processes need to be treated
with coarse approximations in order to predict observable properties of
galaxies.

A detailed analysis of the properties of elliptical galaxies expected in the
framework of hierarchical galaxy formation has been given in a number of papers
\citep{kauffmann96a,baugh96,kauffmann98}.  This early work used an extension of
the Press-Schechter theory to produce Monte Carlo realizations of the merger
trees of dark matter haloes, thus allowing the progenitors of a dark matter
halo to be followed back in time to arbitrarily high redshifts. A drawback of
this approach is its lack of spatial information on the clustering of galaxies.
Another lies in the inherent inaccuracies of the Press-Schechter formalism.
Recent years have witnessed substantial progress in this regard with the advent
of ``hybrid'' techniques where high-resolution N-body simulations of structure
formation are used to directly measure dark matter merger history trees from
simulations which are then combined with semi-analytic simulations of the
galaxy formation physics \citep{kauffmann99,springel01,mathis02}.  This allows
the spatial and kinematic distribution of model galaxies to be predicted as a
function of redshift, providing a more direct and more powerful comparison
between theoretical predictions and observational results. Also, this approach
eliminates much of the uncertainty introduced by Monte-Carlo prescriptions for
producing mock merging histories.

Most previous semi-analytic studies of the properties of ellipticals were
carried out in the framework of a cosmological model with critical matter
density. This cosmogony has been replaced in recent years by the $\Lambda$CDM
scenario, which has become the \emph{de facto} standard cosmological model,
thanks to its concordance with a variety of observational data, including the
most recent cosmic microwave background measurements, distant supernova
observations, and cosmic shear measurements.  Given that both the cosmological
model and the modelling techniques have changed significantly, it is
interesting to revisit the question of the formation and evolution of
elliptical galaxies.  In this paper, we present the results of applying a
semi-analytic model that tracks dark matter halos and their embedded
substructures to the largest high-resolution simulation of cosmic structure
growth ever carried out.  Here we concentrate on the analysis of the star
formation histories, the ages, and the metallicities of model elliptical
galaxies as a function of galaxy mass and of environment.  In a companion
paper, we will study how the distribution of metallicities and ages depends on
the feedback model and on the chemical enrichment scheme employed, exploring
also the relative contribution of these two in shaping observed scaling
properties like the colour--magnitude relation.

The layout of this paper is as follows. In Section~\ref{sec:simulation}, we
briefly describe the simulation used in our study, and in
Section~\ref{sec:model} we give a concise overview of the semi-analytic model
employed for our analysis.  In Section~\ref{sec:sfh}, we discuss how the star
formation history of model elliptical galaxies depends on the stellar mass of
galaxies and on the environment, while Section~\ref{sec:agemet} discusses the
dependence of ages and metallicities on galaxy mass and environment.  Finally,
in Section~\ref{sec:conclusions}, we summarise and discuss our findings, and
give our conclusions.

\section{The simulation}
\label{sec:simulation}

In this study, we analyse a large collisionless cosmological simulation which
follows $N= 2160^3$ particles of mass $8.6\times10^{8}\,h^{-1}{\rm M}_{\odot}$
within a comoving box of size $500\, h^{-1}$Mpc on a side \citep{springel2005}.
The spatial resolution is $5\, h^{-1}$kpc, available everywhere in the periodic
box.  The cosmological model is a $\Lambda$CDM model with parameters
$\Omega_{\rm m}=0.25$, $\Omega_{\rm b}=0.045$, $h=0.73$, $\Omega_\Lambda=0.75$,
$n=1$, and $\sigma_8=0.9$, where the Hubble constant is parameterised as $H_0 =
100\, h\, {\rm km\, s^{-1} Mpc^{-1}}$.  These cosmological parameters are
consistent with recent determinations from the combined analysis of the 2dFGRS
\citep{colless01} and first year WMAP data \citep{spergel03}.

A remarkable aspect of this \emph{Millennium Simulation} carried out by the
Virgo Consortium\footnote{The Virgo consortium (http://www.virgo.dur.ac.uk/) is
  an international collaboration of astronomers dedicated to large scale
  cosmological simulation.}  is its good mass resolution combined with a very
large particle number, more than 10 billion.  It is the largest high-resolution
simulation of cosmic structure growth carried out so far. This provides
substantial statistical power, sampling the formation history even of rare
objects in a representative fashion. \citet{Gao2005} have exploited this to
show that the clustering of halos of fixed mass {\em does depend} on their
formation redshift, an effect that is weak for massive systems but becomes
progressively stronger towards smaller galaxies. We note that simpler schemes
for modelling the galaxy distribution, based for example on the halo occupation
distribution schemes or on Monte Carlo halo merger trees, do not account for
this effect, highlighting the importance of direct simulations of structure
formation for obtaining fully reliable merger histories.

The simulation has sufficient resolution to track the motion of dark matter
substructures in massive halos, making it possible to follow the orbits of
cluster ellipticals in an unambiguous fashion. During the simulation, $64$ time
slices were saved, together with group catalogues and their embedded
substructures, the latter defined as locally overdense and self-bound
structures, identified with the {\small SUBFIND} algorithm \citep{springel01}.
These group catalogues were then used to construct detailed merging history
trees of all gravitationally self-bound dark matter structures
\citep{springel2005}.  The merger trees describe the assembly of about $20$
million galaxies, and form the basic input needed by the semi-analytic
simulation of the galaxy formation process considered in this study.

\section{The semi-analytic model}
\label{sec:model}

Our technique for grafting the semi--analytic model onto the Millennium
Simulation is similar in spirit to that used by \citet{springel01} and De
Lucia, Kauffmann \& White (2004), but has been updated in a number of important
points. A full description of the model is given by \citet{springel2005} and
\citet{croton05}, but we here give a brief account of those aspects that are
particularly relevant for the present study.

One of the key differences of our approach from traditional semi--analytic
models is that we explicitly follow dark matter haloes even after they are
accreted onto larger systems.  This allows the dynamics of satellite galaxies
residing in the infalling haloes to be properly followed until the parent dark
matter `substructure' is completely destroyed because of tidal truncation and
stripping \citep{delucia04a,gao04a}.  At this point, we estimate a residual
survival time of the satellite galaxy and track its position by means of the
most bound particle of the subhalo identified just before the substructure was
disrupted.  In this work, we consider as genuine substructures those subhalos
that contain at least $20$ self--bound particles.  The parent catalogue of dark
matter haloes, which are analysed for substructures and decomposed accordingly,
is identified with a standard friends--of--friends (FOF) algorithm with a
linking length of $0.2$ in units of the mean particle separation.  In our
model, we assume that only the galaxy located at the position of the most bound
particle of the FOF halo - the `central galaxy' - is fed by radiative cooling
from the surrounding halo, i.e.~genuine satellite galaxies cannot replenish
their reservoir of cold gas.  Gas infall and cooling are modelled as described
in detail in \citet{croton05}.

Hierarchical merging trees form the backbone of our semi-analytic model, and
are built for all the self-bound dark matter halos and subhalos in the
Millennium Simulation using the methods described in \citet{springel2005}: the
descendant in the next time slice of each dark matter (sub)halo is identified
as the (sub)halo that contains the largest number of its most tightly bound
particles.  Individual trees are stored separately in a self-contained fashion,
so that the semi-analytic code can be run for each of these trees sequentially,
instead of having to process the whole galaxy population in one single run for
the entire simulation box.  This makes the computation feasible even on small
workstations and allows for easy parallelisation, such that the computation of
the galaxy properties can be repeated with different physical assumptions in a
matter of hours, if desired.

The descriptions we adopted for modelling the various mixing and exchange
processes occurring between different galactic phases (stars and cold gas, hot
diffuse gas in the dark matter haloes, and intergalactic gas outside virialized
haloes) are essentially the same as those in \citet{delucia04b}.  In the
present study, we use their `ejection slow' feedback model, which has been
shown to reproduce both the observed relation between stellar mass and cold
phase metallicity, and the relation between luminosity and cold gas fraction
for galaxies in the local Universe, as well as the observed decline in baryon
fraction from rich clusters to galaxy groups.  As in \citet{delucia04b}, we use
metallicity-dependent cooling rates and luminosities, and we refer the reader
to the original paper for more details on these implementations.

In the present paper, we adopt new parameterisations of star formation and of
the suppression of cooling flows by central galaxy AGN activity, as introduced
by \citet{croton05}. In the following, we briefly summarise the main
characteristics of these physical prescriptions where they differ from those
used in our previous work.

Following \citet{kauffmann96b} and \citet{croton05}, we assume that the star
formation occurs with a rate given by:
\begin{equation}
\label{eq:sfr}
\psi = \alpha (M_{\rm cold} - M_{\rm crit})/t_{\rm dyn},
\end{equation}
where $M_{\rm cold}$ and $t_{\rm dyn} = R_{\rm disc}/V_{\rm vir}$ are the cold
gas mass and the dynamical time of the galaxy, respectively. The dimensionless
parameter $\alpha$ regulates the efficiency of the conversion of gas into
stars.  Star formation is allowed to occur only if the gas surface density is
larger than a critical value (that is used to obtain $M_{\rm crit}$ in
Eq.~\ref{eq:sfr}) given by:
\begin{equation}
\Sigma_{\rm crit} = 1.2\times 10^7 \,\Big(\frac{{V}_{\rm vir}}{200\,{\rm
    km}\,{\rm s}^{-1}}\Big)\,\Big(\frac{{R}}{10\,{\rm kpc}}\Big)^{-1}\,{\rm
    M}_{\odot}\,{\rm kpc}^{-2} 
\end{equation}
Note that in \citet{delucia04b} we did not assume a surface density threshold
for the star formation but we assumed a dependence of the star formation
efficiency on the circular velocity of the parent galaxy.  This assumption was
responsible for delaying the star formation in small haloes until lower
redshifts, which correctly reproduces the observed trend for increasing gas
fraction at lower luminosities.  As explained in \citet{kauffmann96b}, the
introduction of a surface density threshold also naturally reproduces the
observed trend of the gas fraction as a function of galaxy luminosity, due to
the fact that the gas density always remains close to the critical gas surface
density value. We postpone a more detailed investigation on how the census of
ages and metallicities is influenced by the assumptions on the star formation
and feedback model to a forthcoming paper.  With respect to the trends
presented in this paper, we have verified that the two different assumptions
produce very similar results.

As in \citet{delucia04b}, we assume that bulge formation takes place during
mergers: in the case of a `minor' merger, we transfer the stellar mass of the
merged galaxy to the bulge of the central galaxy and update the photometric
properties of this galaxy. The cold gas of the satellite galaxy is added to the
disk of the central galaxy and a fraction of the combined cold gas from both
galaxies is turned into stars as a result of the merger. Any stars that formed
during the burst are also added to the disk of the central galaxy.  If the mass
ratio of the merging galaxies is larger than $0.3$, we assume that we witness a
`major' merger that gives rise to a more significant starburst and destroys 
the disk of the central galaxy completely, producing a purely spheroidal 
stellar distribution. Note that the galaxy can grow a new disc later on, 
provided it is fed by an appreciable cooling flow.  Our starburst 
implementation is based on the `collisional starburst' model introduced by 
\citet{som01} (see \citet{croton05} for details).

Following \citet{croton05}, we extend the spheroid formation by assuming that
bulges can also grow from disk instabilities, based on the the analytic model
for disk formation by \citet*{mo1998}.  We note that the addition of this
chanel for bulge formation does not substantially modify the trends presented
in this paper.  However, this additional physical mechanism changes the
relative fractions of different morphological types.  For the model explored in
this paper, the final fractions of ellipticals, spirals, and lenticulars
brighter than $-18$ in the V--band are $17$, $65$, and $18$ per cent,
respectively, and are very close to the observed relative fractions $13$, $67$,
and $20$ per cent measured by \citet{loveday96}.  If bulge growth through disk
instabilities is switched off, the above fractions become: $7$, $84$, and $8$
per cent respectively.  Considering only galaxies brighter than $-20$, the
fractions cited above become: $23$, $58$, and $19$ for our default model and
$13$, $67$, and $20$ for a model where bulge growth through accretion is
switched off.  These numbers suggest that this channel of bulge formation may
be more important for fainter ellipticals.  We will come back to this issue in
Sec.~\ref{sec:agemet}.

As in previous work, we determine the morphology of our model galaxies by using
the B--band bulge--to--disc ratio together with the observational relation by
\citet{simien} between this quantity and the galaxy morphological type.  For
the numbers quoted above and for the following analysis, we classify as
ellipticals all galaxies with $\Delta M < 0.4$ ($\Delta M= M_{\rm bulge} -
M_{\rm total}$), as spirals or irregulars all galaxies with $\Delta M > 1.56$,
and as lenticulars (S0) all galaxies with intermediate value of $\Delta M$.  We
include in our analysis all galaxies with stellar mass larger than
$3\times10^8\,{\rm M}_{\odot}$.  Note that, although we are essentially
complete down to this mass limit, the less massive galaxies included in this
analysis reside in substructures whose mass accretion history can be followed
back in time in many cases only for a small number of snapshots.  The
determination of the morphological type then becomes quite noisy and uncertain.
We estimate that our morphological type determination is robust for galaxies
with mass equal to a few times $10^9\,{\rm M}_{\odot}$. The inclusion of
galaxies below this limit, however, does not affect our main results of this
study.  When necessary, we will explicitly show results including only
galaxies with mass larger than $4\times10^9\,{\rm M}_{\odot}$, whose
morphological type can be considered `secure'.  Our final sample contains
$1,031,049$ elliptical galaxies with stellar mass larger than
$4\times10^9\,{\rm M}_{\odot}$.  $810,486$ of these have stellar mass larger
than $1\times10^{10}\,{\rm M}_{\odot}$.

Finally, we use the model of \citet{croton05} to describe central heating by
AGN in massive groups and clusters and the associated suppression of cooling
flows.  In this model, gas condensation in massive systems is efficiently
suppressed by `radio mode' outflows that occur when a massive black hole finds
itself at the centre of a static hot gas halo.  The importance of these
outflows grows with decreasing redshift and with the mass of the system.  We
refer to the original paper for full details and the physical motivation for
this feedback implementation.  We will later discuss in more detail the effect
that this particular implementation has on the trends presented in our study.
\section{The star formation history of elliptical galaxies}
\label{sec:sfh}

\begin{figure}
\bc
\hspace{-1.4cm}
\resizebox{8cm}{!}{\includegraphics{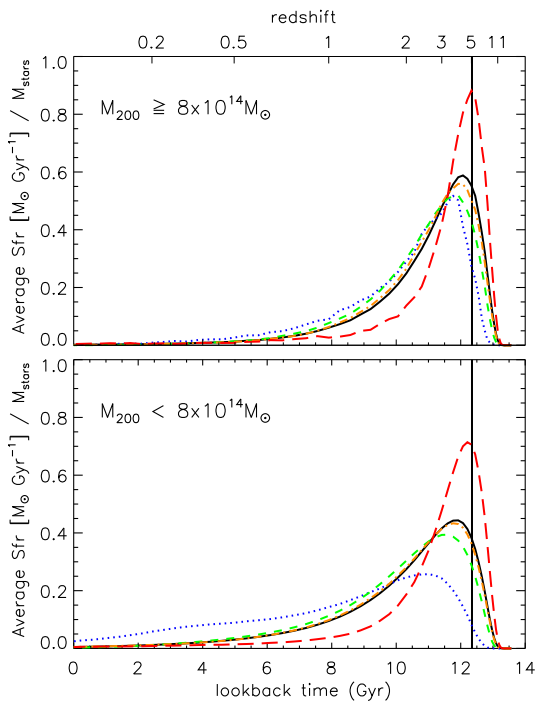}}\\%
\caption{Average star formation histories of model elliptical galaxies
  split into bins of different stellar mass, normalised to the total mass of
  stars formed.  The two panels are for galaxies residing in haloes of
  different mass, as indicated by the labels.  In both panels, the solid line
  shows the average star formation history for all the elliptical galaxies in
  the sample under investigation.  The long dashed, dash-dotted, dashed, and
  dotted lines refer to galaxies with stellar mass $\simeq\,10^{12}$,
  $10^{11}$, $10^{10}$, and $10^{9}\,{\rm M}_{\sun}$ respectively.  The
  vertical line in both panels is included to guide the eye.}
\label{fig:sfr}
\ec
\end{figure}

As discussed earlier, the uniformly red and old stellar populations of
elliptical galaxies have traditionally been interpreted as evidence for a
formation scenario in which these galaxies form in a single intense burst of
star formation at high redshift and then passively evolve to the present day.
Direct observations of the formation and evolution of early type galaxies are,
however, difficult, and are plagued by the so called `progenitor-bias'
\citep{dokkumfranx1996}.  One approach that is adopted to constrain the
formation mechanism of these galaxies is that of studying in detail their
stellar populations by means of population synthesis techniques. This method
has its roots in a pioneering study by \citet{tinsley1972}, and has become
increasingly popular after the introduction of more detailed population
synthesis models \citep{bruzual1983,guiderdoni1987,buzzoni1989}.  Recent
improvements have come through the development of medium to high resolution
spectral models that include quite complete libraries of stellar spectra and
improved treatments of stellar evolutionary theory
\citep*{vazdekis2001,BC2003,TMB2003}.

The use of these more sophisticated models, together with the acquisition of
better and larger amounts of data have recently established firm evidence for a
mass--dependent evolutionary history of the elliptical galaxy population
\citep{delucia04c,kodama2004,thomas04,vandervel05,treu05}. The data suggest
that less massive ellipticals have more extended star formation histories than
their more massive counterparts, giving them a {\it lower} characteristic {\it
  formation redshift}, in marked contrast to naive expectations based on the
growth of dark matter halos in hierarchical CDM cosmologies.  This
observational finding is compatible with the ``down--sizing'' scenario for star
formation proposed earlier \citep{faber1995,cowie96}.

\begin{figure*}
\bc
\resizebox{18cm}{!}{\includegraphics{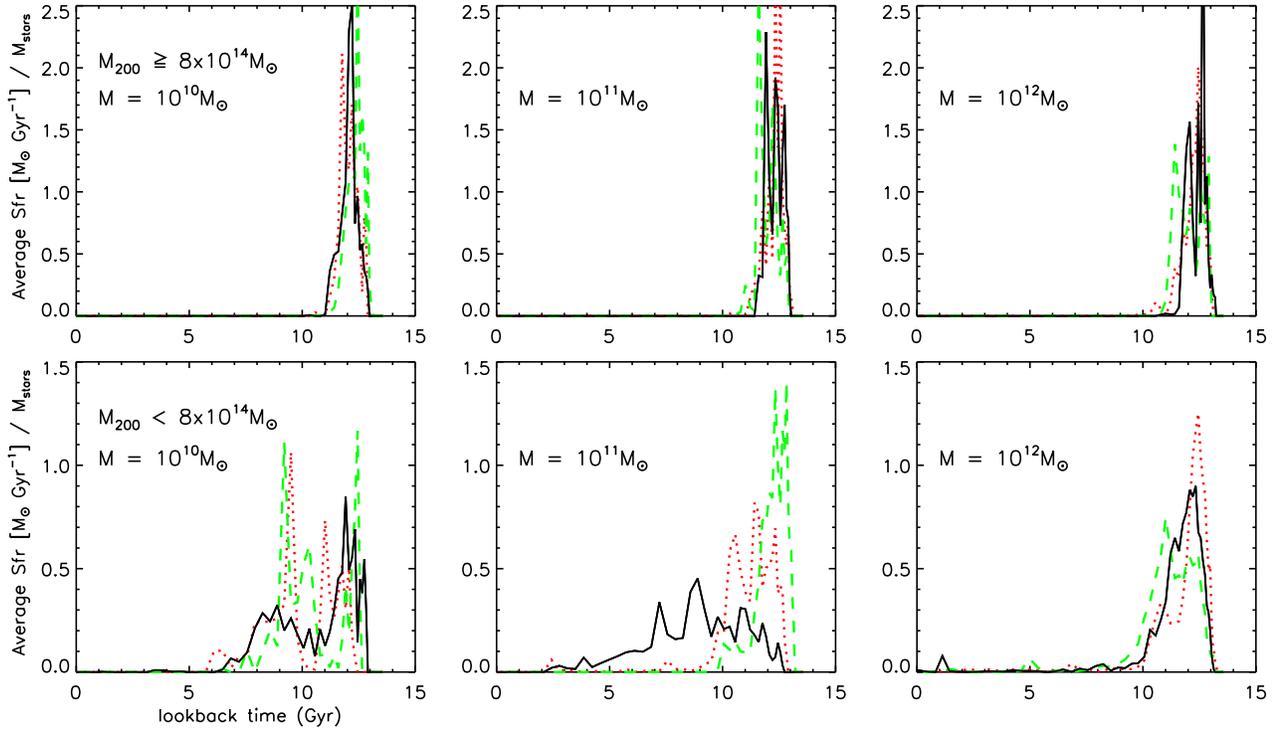}}\\%
\caption{Each panel shows the star formation history for three randomly
  selected model elliptical galaxies in the given mass bin.} 
\label{fig:sfronefile}
\ec
\end{figure*}
In this section, we study in detail the star formation histories of model
elliptical galaxies and their dependence on stellar mass and environment.  In
Fig.~\ref{fig:sfr}, we show the average star formation history, normalised to
the total mass of stars formed, for model elliptical galaxies split into
different bins of stellar mass. The two panels are for galaxies residing in
haloes of different mass.  In the following we will refer to galaxies in haloes
with mass $\gtrsim 8\times10^{14}\,{\rm M}_{\sun}$ as `cluster' ellipticals and
as `field' ellipticals to all model ellipticals residing in less massive
haloes.  In both panels, the solid line shows the average star formation
history for all the elliptical galaxies in the sample under investigation.  The
long dashed, dash-dotted, dashed, and dotted lines refer to galaxies with
stellar mass $\simeq 1\times10^{12}$, $1\times10^{11}$, $1\times10^{10}$, and
$1\times10^{9}\,{\rm M}_{\sun}$, respectively\footnote{The actual bin size used
  for Figs.~\ref{fig:sfr} and \ref{fig:sfronefile} is $7\times10^{x}\,{\rm
    M}_{\sun} < {\rm M}_{\rm stars} < 2\times10^{x+1}\,{\rm M}_{\sun}$, with
  $x$ = $8$, $9$, $10$, and $11$.}.  The vertical lines are included to guide
the eye and mark the peak of the $10^{12}\,{\rm M}_\odot$-ellipticals in the
top panel.

Fig.~\ref{fig:sfr} shows the most important result of this paper: more massive
elliptical galaxies have star formation histories that peak at higher redshifts
($\simeq 5$) than lower mass systems, and can reach star formation rates up to
several thousands of solar masses per year for galaxies ending up in overdense
regions.  Less massive elliptical galaxies have star formation histories that
peak at progressively lower redshifts and are extended over a longer time
interval.

A comparison of the top and bottom panels of Fig.~\ref{fig:sfr} shows that the
qualitative behaviour for `field' and `cluster' ellipticals is the same, but
that for fixed mass, the star formation histories of field ellipticals are
predicted to be more extended than those of ellipticals in clusters.  This is a
natural outcome of the hierarchical scenario, where haloes in regions of the
Universe that are destined to form a cluster collapse earlier and merge more
rapidly.  The star formation histories shown in Fig.~\ref{fig:sfr} represent
averages computed over all the elliptical galaxies in the simulation box, but
the trends remain true also when a much smaller volume of the simulation, and
hence a much smaller sample size, is analysed.  Fig.~\ref{fig:sfronefile}
shows the star formation histories of randomly selected elliptical galaxies in
different mass bins and in different environments.  The figure shows that
individual star formation histories display a much more `bursty' behaviour than
those shown in Fig.~\ref{fig:sfr}.  This reflects our assumption that bulge
formation takes place during merger-induced bursts, which naturally gives the
star formation histories of individual systems a bursty nature quite different
from the smooth history seen for the population average.  We will comment more
on the implications of this for the scatter of the ages for the model
elliptical galaxies in the following section.

\begin{figure}
\bc
\hspace{-1.4cm}
\resizebox{8cm}{!}{\includegraphics{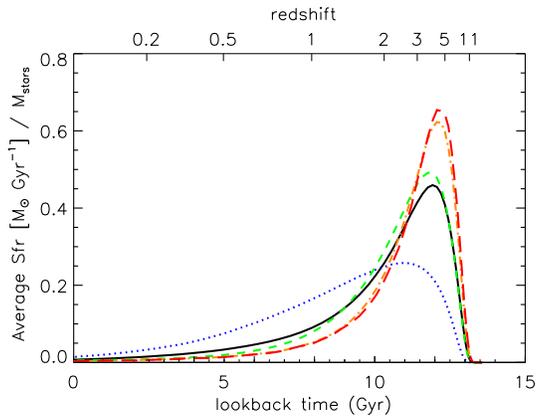}}\\%
\caption{As in Fig.~\ref{fig:sfr}, but with model elliptical galaxies split
  into bins of different {\em parent halo mass}.  The solid line shows the
  average mass--weighted star formation history for all the elliptical galaxies
  in the sample under investigation.  The long dashed, dash-dotted, dashed, and
  dotted lines refer to galaxies residing in haloes with mass ${\rm
    M}_{200}\simeq 10^{15}$, $10^{14}$, $10^{13}$, and $10^{12}\,{\rm
    M}_{\sun}$, respectively.}
\label{fig:sfrhalo}
\ec
\end{figure}

In Fig.~\ref{fig:sfrhalo}, we show the star formation histories again, but this
time split into bins of different parent {\em halo mass}. The long dashed,
dash-dotted, dashed, and dotted lines are for elliptical galaxies in haloes of
mass $\simeq 1\times10^{15}$, $1\times10^{14}$, $1\times10^{13}$, and
$1\times10^{12}\,{\rm M}_{\sun}$ respectively\footnote{The actual bin size used
  is $7\times10^{x}\,{\rm M}_{\sun} < {\rm M}_{\rm 200} < 2\times10^{x+1}\,{\rm
    M}_{\sun}$, with $x$ = $11$, $12$, $13$, and $14$.}.  Only galaxies with
stellar mass larger than $4\times10^{9}\,{\rm M}_{\odot}$ are used here. The
solid line shows the average mass--weighted star formation history for all the
galaxies in the sample. The faster evolution of proto--cluster regions produces
star formation histories that peak at higher redshifts for galaxies in more
massive haloes.  Given that galaxies of a fixed stellar mass occur in haloes
covering a wide range of masses, it is not surprising that the dependence of
the star formation history on halo mass is much weaker than that on galaxy
stellar mass.

\section{The distribution of ages and metallicity}
\label{sec:agemet}

\begin{figure}
\bc
\resizebox{8cm}{!}{\includegraphics{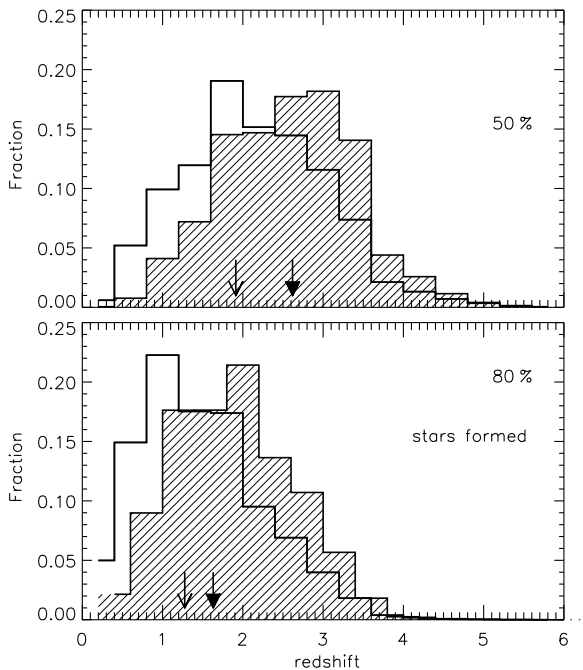}}\\%
\caption{Distribution of the formation redshifts of model elliptical
  galaxies. In the upper (lower) panel, the formation redshift is defined as
  the redshift when $50$ per cent ($80$ per cent) of the stars that make up the
  elliptical galaxy at redshift $z=0$ are already formed.  The shaded histogram
  is for elliptical galaxies with stellar mass larger than $10^{11}\,{\rm
    M}_{\sun}$, while the open histogram is for all the galaxies with mass
  larger than $4\times10^{9}\,{\rm M}_{\sun}$.  Arrows indicate the medians of
  the distributions, with the thick arrows referring to the shaded histograms.
  Note that more massive ellipticals typically form their stars earlier.}
\label{fig:formtime}
\ec
\end{figure}
We now turn to an analysis of the distribution of ages and metallicities of
model elliptical galaxies as a function of stellar mass and environment.  In
Fig.~\ref{fig:formtime}, we show the distribution of the formation redshifts
for model elliptical galaxies.  We define the formation redshift as the
redshift when $50$ per cent (or $80$ per cent) of the stars that make up the
final elliptical galaxy at redshift zero are already formed.  The shaded
histograms are for model elliptical galaxies with stellar mass larger than
$10^{11}\,{\rm M}_{\sun}$, while the open histogram is for all galaxies with
secure morphology (stellar mass larger than $4\times10^{9}\,{\rm M}_{\sun}$).
The figure clearly demonstrates that the stars in more massive ellipticals are
on average older than those in their less massive counterparts, but the scatter
of the distribution is rather large and there is a non-negligible fraction of
model galaxies whose stars are formed relatively late.

\begin{figure}
\bc
\resizebox{8cm}{!}{\includegraphics{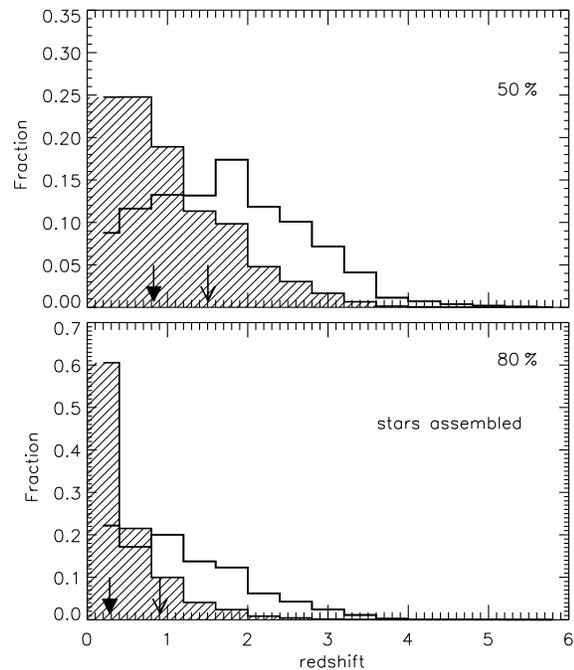}}\\%
\caption{As in Fig.~\ref{fig:formtime}, but for the assembly redshifts
  of model elliptical galaxies. We define the assembly redshift as the time
  when $50$ per cent ($80$ per cent) of the stars that make up the galaxy at
  redshift zero are already assembled in one single object.  Note that more
  massive ellipticals typically assemble their stars later (cf
  Fig.~\ref{fig:formtime}).}
\label{fig:asstime}
\ec
\end{figure}

It is important, however, to distinguish the early {\em formation times} of the
stars that make up the elliptical galaxy population (reflected in the
`down-sizing' scenario) from the {\em assembly time} of the more massive
ellipticals. If massive ellipticals form a large fraction of their stars in a
number of distinct progenitor systems before they coalesce, these two times may
well be quite different. Fig.~\ref{fig:asstime} demonstrates that this is
indeed the case in our model.  We here show the distribution of the assembly
redshifts for the same galaxies that we analysed in Fig.~\ref{fig:formtime}.
We define the assembly time as the redshift when $50$ per cent (or $80$ per
cent) of the final stellar mass is already contained in a single object.  For
galaxies more massive than $10^{11}\,{\rm M}_{\sun}$, the median redshift when
half of the stars are formed is $\sim 2.5$ (upper panel of
Fig.~\ref{fig:formtime}), but for the same galaxies, half of their stars are
typically assembled in a single object only at redshift $\sim 0.8$ (upper panel
of Fig.~\ref{fig:asstime}).  In addition, more massive galaxies assemble later
than less massive ones, and only about half of the model elliptical galaxies
have a progenitor with mass at least equal to half of their final mass at
redshifts $\gtrsim 1.5$. The assembly history of ellipticals hence parallels
the hierarchical growth of dark matter halos, in contrast to the formation
history of the stars themselves.  Note that the `gap' between assembly
redshifts and formation redshifts for the stars grows towards more massive
ellipticals.  Figs.~\ref{fig:formtime} and \ref{fig:asstime} imply that a
significant fraction of present elliptical galaxies has assembled relatively
recently through purely stellar mergers.  This finding agrees with recent
observational results \citep{dokkum05,faber05,tran05,bell05}.

Table \ref{tab:tab1} lists, for different mass bins, medians and upper and
lower quartiles of the distributions of lookback times corresponding to the
formation and assembly redshifts defined above.

\begin{table*}
\caption{Formation times and assembly times for model elliptical galaxies in
  different mass bins.  The first two columns indicate the extrema of the mass 
  bins. The next columns list the lower quartile, the median, and upper
  quartile for each of the formation and assembly times defined in the text. 
  Tf$50$, Tf$80$ represent the lookback times corresponding to the redshifts
  when $50$ or $80$ per cent of the stars were first formed. Ta$50$ and Ta$80$
  represent the lookback times corresponding to the redshifts when $50$ or $80$
  per cent of the mass was first assembled in a single object.  All times are
  in Gyr. Masses are in units of ${\rm M}_{\odot}$.} 

\begin{tabular}{ccrclrclrcrrcl}
\hline
${\rm M}_{\rm low}$ & ${\rm M}_{\rm up}$ & & \hspace{-0.15cm}Tf$50$ & & &\hspace{-0.18cm}Tf$80$ & & &\hspace{-0.18cm}Ta$50$ & & & \hspace{-0.18cm}Ta$80$ \\
\hline
$2.5\times10^{9}$   & $1.25\times10^{10}$ &$8.90 $&\hspace{-0.25cm}$10.06$ & \hspace{-0.25cm}$11.01$ &\hspace{0.25cm}$6.84$ &\hspace{-0.25cm}$8.58 $ &\hspace{-0.25cm}$10.06$&\hspace{0.25cm}$8.25$&\hspace{-0.25cm}$9.78$&\hspace{-0.25cm}$10.79$&\hspace{0.25cm}$6.10$&\hspace{-0.25cm}$8.25$&\hspace{-0.25cm}$9.78$\\
$1.25\times10^{10}$ & $6.25\times10^{10}$ &$9.20 $&\hspace{-0.25cm}$10.31$ & \hspace{-0.25cm}$11.01$ &\hspace{0.25cm}$7.20$ &\hspace{-0.25cm}$8.90 $ &\hspace{-0.25cm}$10.06$&\hspace{0.25cm}$7.91$&\hspace{-0.25cm}$9.78$&\hspace{-0.25cm}$10.56$&\hspace{0.25cm}$5.35$&\hspace{-0.25cm}$7.91$&\hspace{-0.25cm}$9.50$\\ 
$6.25\times10^{10}$ & $3.12\times10^{11}$ &$9.78 $&\hspace{-0.25cm}$10.56$ & \hspace{-0.25cm}$11.22$ &\hspace{0.25cm}$7.56$ &\hspace{-0.25cm}$9.20 $ &\hspace{-0.25cm}$10.31$&\hspace{0.25cm}$4.97$&\hspace{-0.25cm}$7.91$&\hspace{-0.25cm}$9.78 $&\hspace{0.25cm}$1.76$&\hspace{-0.25cm}$4.22$&\hspace{-0.25cm}$7.20$\\ 
$3.12\times10^{11}$ & $1.56\times10^{12}$ &$11.01$&\hspace{-0.25cm}$11.41$ & \hspace{-0.25cm}$11.76$ &\hspace{0.25cm}$9.50$ &\hspace{-0.25cm}$10.31$ &\hspace{-0.25cm}$10.79$&\hspace{0.25cm}$3.13$&\hspace{-0.25cm}$4.97$&\hspace{-0.25cm}$6.84 $&\hspace{0.25cm}$0.83$&\hspace{-0.25cm}$2.09$&\hspace{-0.25cm}$3.85$\\ 
\hline
\end{tabular}
\label{tab:tab1}
\end{table*}

\begin{figure}
\bc
\hspace{1cm}
\resizebox{9cm}{!}{\includegraphics{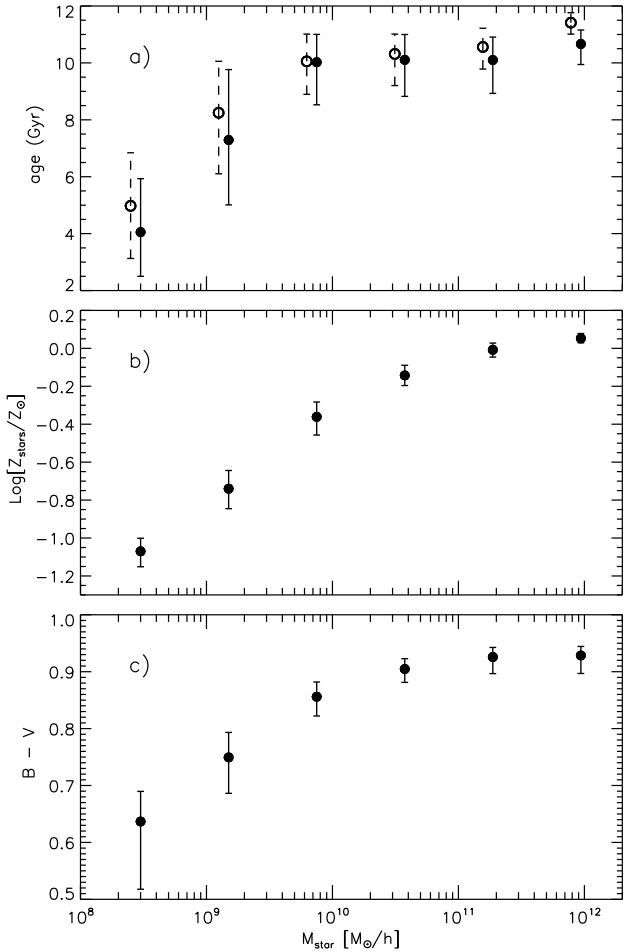}}\\%
\caption{Median luminosity--weighted age (panel a), stellar metallicity
  (panel b), and B$-$V colour (panel b) of model elliptical galaxies as a 
  function of their stellar mass.  Symbols indicate the median value of the
  distributions at each mass, while the error bars link the upper and lower
  quartiles.  The open symbols in the top panel correspond to the lookback time
  of the upper panel in Fig.~\ref{fig:formtime}.} 
\label{fig:ell_mass}
\ec
\end{figure}

The large volume of our simulation allows us to study how properties of model
elliptical galaxies depend on their stellar mass and on the environment.
Fig.~\ref{fig:ell_mass} shows how the luminosity--weighted age (panel a), the
metallicity of the stellar component (panel b), and the B$-$V colour (panel c),
depend on the galaxy stellar mass.  In each panel, filled circles represent the
median of the distributions, while the error bars mark the upper and lower
quartiles.  In the upper panel, the empty circles show the lookback times
corresponding to the formation redshifts as defined in the upper panel of
Fig.~\ref{fig:formtime}.
The ages of model elliptical galaxies range from $\simeq 4\,{\rm Gyr}$ for
galaxies with mass a few times larger than $10^8\,{\rm M}_{\odot}$ to $\simeq
10\,{\rm Gyr}$ for galaxies with mass $\simeq 10^{12}\,{\rm M}_{\odot}$.  It is
interesting to note that the lookback time corresponding to the redshift when
half of the stars had formed is a very good approximation to the
luminosity--weighted age over the full range of masses shown.  The age of model
elliptical galaxies also seems to flatten at stellar masses $\simeq
10^{10}\,{\rm M}_{\odot}$.  The same flattening is observed for the B$-$V
colour and, although weaker, for the stellar metallicity.  Note that the
scatter in these quantities (particularly for the colour and metallicity) is
very small, indicating that the main driver of these trends is the stellar
mass, as also reflected in Fig.~\ref{fig:sfr}.  We note in passing that our
model elliptical galaxies follow a colour-magnitude relation that is well
defined up to $z\simeq1$.  The results shown in Fig.~\ref{fig:ell_mass} also
indicate that this relation is mainly driven by metallicity, in line with the
common interpretation of the evolution of the slope and the zero-point of the
observed relation as a function of redshift.  We plan to investigate the
relative contribution of age and metallicity in shaping the observed
colour-magnitude relation in a future paper.

\begin{figure}
\bc
\hspace{1cm}
\resizebox{9cm}{!}{\includegraphics{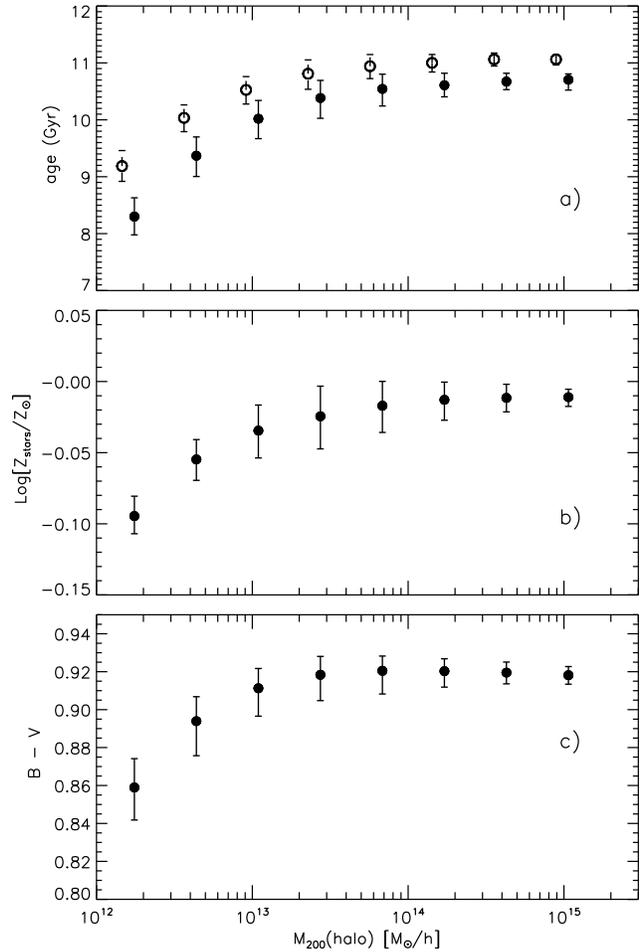}}\\%
\caption{The same quantities as in Fig.~\ref{fig:ell_mass}, but now calculated
  as mass-weighted average over all galaxies in a halo and shown as a function
  of the virial mass of the halo. Only elliptical galaxies with stellar mass
  larger than $4\times10^{9}\,{\rm M}_{\odot}$ are included here.}
\label{fig:ell_halomass}
\ec
\end{figure}

In Fig.~\ref{fig:ell_halomass} we show the same quantities as in
Fig.~\ref{fig:ell_mass} but as a function of the virial mass of the halo in
which the galaxies reside.  Only galaxies with stellar mass larger than
$4\times10^{9}\,{\rm M}_{\odot}$ are included here.  For each halo, we compute
the mass--weighted average age, metallicity and colour.
Fig.~\ref{fig:ell_halomass} shows the median of these values and the upper and
lower quartiles of the distribution.  Elliptical galaxies in high density
environments are on average older, more metal rich, and redder than isolated
elliptical galaxies.  Elliptical galaxies in clusters form a homogeneously old
population and elliptical galaxies in groups (haloes with mass $\simeq
10^{13}\,{\rm M}_{\sun}$) are as old as or slightly younger than galaxies in
massive clusters (haloes with mass $\simeq 10^{15}\,{\rm M}_{\sun}$), while
elliptical galaxies in smaller groups exhibit a lower median age.  This is in
agreement with recent observational results \citep{terlevich02,proctor04}.  The
median stellar metallicities of galaxies in our clusters are higher than the
corresponding values for the field.  Elliptical galaxies in these systems show
a remarkably small spread in colour with a median B$-$V colour that is almost
independent of environment down to the mass-scale $\simeq 10^{13}\,{\rm
  M}_{\sun}$ of groups.
 
\begin{figure}
\bc
\hspace{1cm}
\resizebox{9cm}{!}{\includegraphics{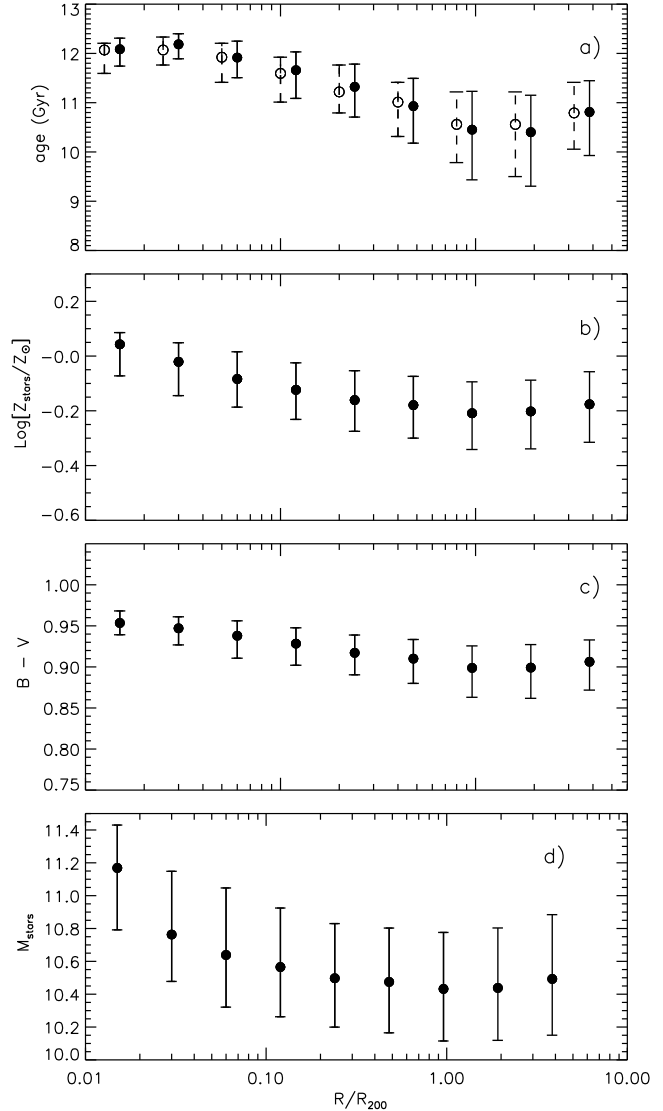}}\\%
\caption{Median luminosity--weighted age (panel a), stellar metallicity
  (panel b), B$-$V colour (panel c), and stellar mass (panel d) for model
  elliptical galaxies in dark matter haloes with mass $\gtrsim
  8\times10^{14}\,{\rm M}_{\sun}$ as a function of the distance from the
  cluster centre.  Symbols and lines have the same meaning as in
  Fig.~\ref{fig:ell_mass}.}
\label{fig:ell_cluster}
\ec
\end{figure}

We can also investigate how the properties of model elliptical galaxies depend
on cluster--centric distance.  This is shown in Fig.~\ref{fig:ell_cluster},
based on all elliptical galaxies in haloes with ${\rm M}_{200} \gtrsim
8\times10^{14}\,{\rm M}_{\sun}$ and with stellar mass larger than
$4\times10^{9}\,{\rm M}_{\sun}$.  We find in total $51$ clusters in the whole
simulation box with virial mass larger than this value.
Fig.~\ref{fig:ell_cluster} shows that galaxies closer to the centre are on
average older and more metal rich than galaxies at the outskirts of these
clusters.  The bottom panel of Fig.~\ref{fig:ell_cluster} shows that this
trend is partly driven by mass segregation.  A radial dependence of galaxy
properties is, however, also a natural consequence of the fact that mixing of
the galaxy population is incomplete during cluster assembly. This implies that
the cluster--centric distance of the galaxies is correlated with the time they
were accreted onto a larger structure \citep{diaferio2001,gao04a}.  The values
of the median age, metallicity, and colour all flatten at distance $\simeq
\rm{R}_{200}$.  The scatter shown in Fig.~\ref{fig:ell_cluster} is rather large
but in the very centre of the clusters, where elliptical galaxies are all very
old ($\simeq 12\,{\rm Gyr}$), they have about solar metallicity, and have very
red B$-$V colour ($\simeq 0.95$).

In the hierarchical galaxy formation scenario, elliptical galaxies form through
mergers of smaller units, and larger systems are expected to be made up by a
larger number of progenitor galaxies.  A very interesting question is therefore
how large the number of progenitor systems of galaxies is, and how this number
varies as a function of final mass.  To get a quantitative handle on this
question, we define for each galaxy an {\it effective number of stellar
  progenitors} by computing the quantity
\begin{equation}
{N}_{\rm eff} = \frac{M_{\rm final}^2} {2\,\sum_i  m_i\, M_{i, \rm
    form}},
\label{eq:neff} 
\end{equation}
where $m_i$ denotes the masses of all the stars that make up a galaxy of final
mass $M_{\rm final} = \sum_i m_i$. The quantity $M_{i, \rm form}$ gives the
stellar mass of the galaxy within which the star $i$ formed, at the time of
formation of the star.  In the case where all stars form in a single object
that grows to the final stellar mass without experiencing any merger,
Eq.~(\ref{eq:neff}) can be viewed as a discretised form of the integral
\begin{displaymath}
N_{\rm eff} = \frac{M_{\rm final}^2}{\int_0^{M_{\rm final}}2\,M\,{\rm d}M},
\end{displaymath}
which evaluates to $N_{\rm eff}= 1$ independent of the detailed star formation
history. However, if a galaxy is assembled from several pieces we expect a
larger value of $N_{\rm eff}$, because then the values of $M_{i, \rm form}$
that enter the sum in the denominator of Eq.~(\ref{eq:neff}) become lower.  For
example, if the stars of a galaxy were formed in two progenitors of equal final
size, which then merged into a single object without any further star
formation, we obtain $N_{\rm eff}= 2$.  Note that in more general cases we will
obtain fractional values for $N_{\rm eff}$ due to the mass-weighting of the
progenitors, which is built into the definition of $N_{\rm eff}$.  For example,
if a galaxy is made up of three pieces that contain one half, one quarter and
one quarter of the final stellar mass, respectively, one gets $N_{\rm eff} =
8/3$, which is less than the absolute number of progenitors. This reflects the
fact that the majority of the stars formed in a single object.  The
mass-weighting hence delivers an effective count of the progenitors by giving
weight only to those progenitor systems that contribute significantly to the
final stellar mass of the galaxy. In contrast, a count of the total number of
progenitors would be dominated by the large number of negligibly small
satellites that fall into a galaxy during its growth in a hierarchical
universe.  We therefore argue that $N_{\rm eff}$ is a more useful proxy for the
number of significant mergers required to assemble a galaxy.  We caution
however that a straightforward interpretation of $N_{\rm eff}$ in the context
of spheroid formation is complicated by the fact that bulges can also grow in
our model without mergers, through disk instabilities.

\begin{figure}
  \bc \hspace{1cm}
  \resizebox{9cm}{!}{\includegraphics{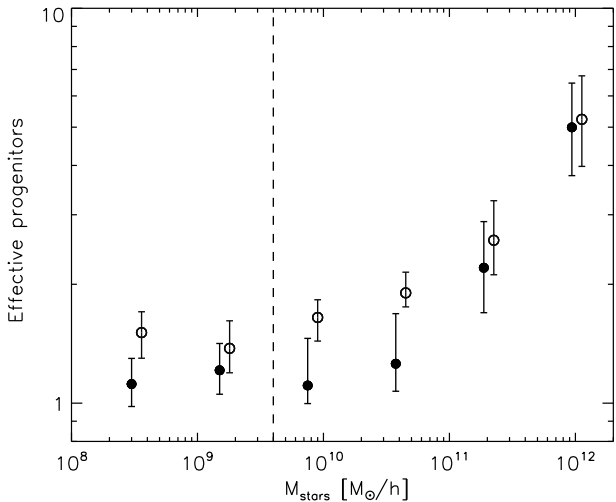}}\\%
\caption{Effective number $N_{\rm eff}$ of progenitors as a
  function of galaxy stellar mass.  Filled circles represent the median of
  the distribution in our default model, while the error bars indicate the
  upper and lower quartiles.  Empty circles and the corresponding error
  bars are for a model where bulge formation through disk instability is
  switched off.  The vertical dashed line corresponds to the limit above which
  our morphological type determination is robust (see Sec.~\ref{sec:model}).}
\label{fig:mergers}
\ec
\end{figure}

In Fig.~\ref{fig:mergers}, we show the effective number of progenitors as a
function of galaxy stellar mass.  Filled circles represent the median of the
distributions in our default model while empty circles represent the median of
the distributions in a model where bulge growth through disk instability is
switched off.  Interestingly, bulge growth through disk instability seems to be
an efficient process for intermediate mass ellipticals but rather ineffective
for the most massive ellipticals in our sample.  As expected, more massive
galaxies are made up of more pieces.  Fig.~\ref{fig:mergers} shows that the
number of effective progenitors is less than $2$ up to stellar masses of
$\simeq 10^{11}\,{\rm M}_{\sun}$, indicating that the formation of these
systems typically involves only a small number of major mergers.  Only galaxies
more massive than $\simeq 10^{11}\,{\rm M}_{\sun}$ are built up through a
larger number of mergers, reaching up to $N_{\rm eff}\simeq 5$
for the most massive galaxies.  We recall that these most massive elliptical
galaxies are, however, also the ones with the oldest stellar populations. We
note that the monolithic collapse scenario would predict $N_{\rm eff}=1$ for
these large ellipticals, in marked difference to our hierarchical prediction.

\section{Discussion and Conclusions}
\label{sec:conclusions}

We have combined a large high-resolution cosmological $N$--body simulation with
semi--analytic techniques to investigate the formation and evolution of
elliptical galaxies in a hierarchical merger model.  Understanding the
formation and the evolution of these systems represents an issue of fundamental
interest as $50$ per cent or more of the stellar mass in the local Universe
appears to be in early--type systems and bulges \citep{bell03}.

In this paper, we have focused on the dependence of the star formation
histories, ages, and metallicities on environment and on galaxy stellar mass.
In our model, we find that elliptical galaxies in denser environments are on
average older, more metal rich, and redder than the general population of
`field' ellipticals.  This can be attributed to the fact that high density
regions form from the highest density peaks in the primordial field of density
fluctuations, whose evolution is somewhat accelerated with respect to regions
of `average' density.  There is also a clear trend for increasing ages and
metallicities, and for redder colours, with decreasing cluster--centric
distance.  This can again be viewed as a natural expectation of hierarchical
models where the distance of the galaxies from the cluster centre is correlated
with the time they were accreted onto the larger system.  When this infall
happens, we assume that the galaxy is stripped of its hot gas reservoir so it
is no longer able to accrete fresh material for star formation.  The galaxy
then rapidly consumes its cold gas moving towards the red sequence.

We have also investigated how the properties of model elliptical galaxies
change as a function of the stellar mass.  We have shown -- and this is perhaps
the most important result of our study -- that in our model the most massive
elliptical galaxies have the oldest and most metal rich stellar populations, in
agreement with observational results \citep[see for example][]{nelan05}.  In
addition, they are also characterised by the shortest formation time--scales,
in qualitative agreement with the recently established down-sizing scenario
\citep{cowie96}. However, these old ages are in marked contrast to the late
assembly times we find for these galaxies. In fact, our results show that
massive ellipticals are predicted to be assembled {\em later} than their lower
mass counterparts, and that they have a larger effective number of progenitor
systems. This is a key difference between the hierarchical scenario and the
traditional monolithic collapse picture.

\begin{figure}
\vspace{-.8cm}
\bc
\hspace{-1.cm}
\vspace{-0.5cm}
\resizebox{9cm}{!}{\includegraphics{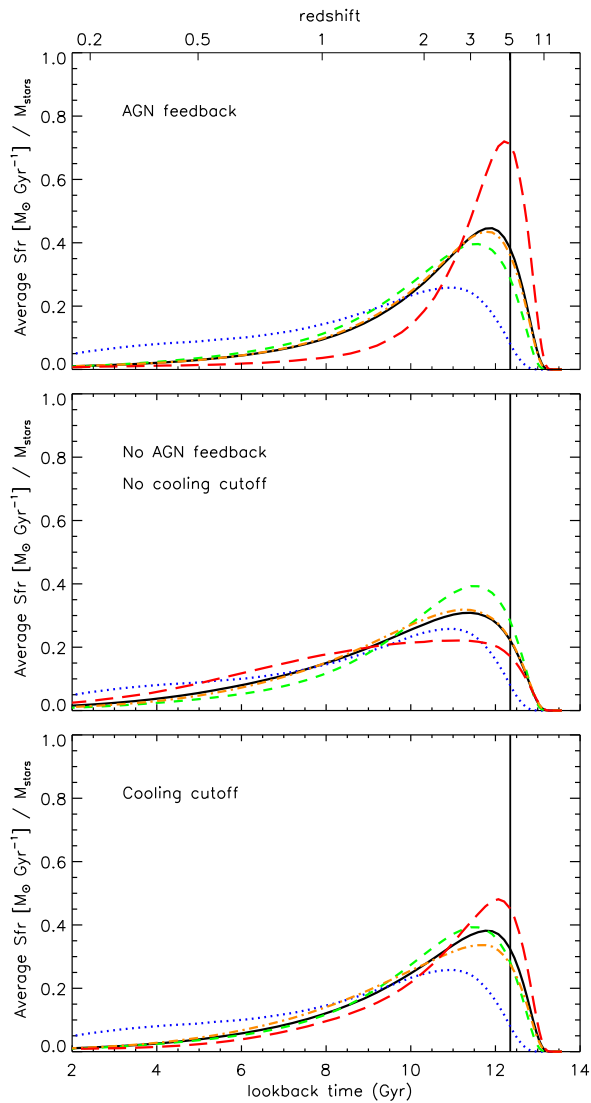}}\\%
\caption{Average star formation histories of model elliptical galaxies
  split into bins of different stellar mass, normalised to the total mass of
  stars formed.  In the upper panel, results are shown for the `standard' model
  used for our analysis, which employs the suppression of cooling flows by
  central galaxy AGN activity, as introduced by \citet{croton05}.  In the
  middle panel, results are shown for a model without AGN feedback and without
  any cooling cutoff.  Finally, in the bottom panel, results are shown for a
  model without AGN feedback but with a cooling cutoff with a critical velocity
  equal to $350\,{\rm km}\,{\rm s}^{-1}$.  Different line styles have the same
  meaning as in Fig.~\ref{fig:sfr}.}
\label{fig:sfrcheck}
\ec
\end{figure}

Our results disagree with previous semi-analytic models that found a trend for
more massive ellipticals to be {\it younger} than less massive ones
\citep{kauffmann96a,baugh96,kauffmann98}.  In order to understand the origin of
this discrepancy we have re-run our model with different assumptions.
Fig.~\ref{fig:sfrcheck} shows the average star formation histories of all model
elliptical galaxies split into bins of different stellar mass and normalised to
the total mass of stars formed, as in Fig.~\ref{fig:sfr}.  In the upper panel,
we repeat the results obtained for the `standard' model used in our analysis,
which employs the suppression of cooling flows by central AGN activity as
introduced by \citet{croton05}.  In the middle panel, we show the same results
but for a model in which no AGN feedback and no artificial cooling cutoff is
included.  Finally, in the bottom panel, we show results obtained for a model
in which galaxy cooling is switched off in haloes with ${\rm V}_{\rm vir} >
350\,{\rm km}\,{\rm s}^{-1}$ - the {\em ad hoc} suppression used in many
previous models, including those of \citet{delucia04b}.

Fig.~\ref{fig:sfrcheck} clearly shows that when no suppression of the
condensation of gas in massive haloes is considered, the most massive
ellipticals have the most extended star formation histories.  Too many massive systems are, however, produced at redshift zero, at odds with
observations.  Late mergers and late accretion, which still involve a
substantial amount of gas in this model, cause the formation of luminous and
young bulge stars.  An artificial cutoff of the gas condensation, similar to
that employed in previous models, produces results that are qualitatively
similar \citep[see also][]{deluciaPhD} to those obtained with the more
physically motivated AGN model introduced by \citet{croton05}, as is shown in
the bottom panel of Fig.~\ref{fig:sfrcheck}.  The figure also indicates
however, that the model does not produce a monotonic behaviour as a function of
stellar mass.

\begin{figure}
  \bc \hspace{1cm}
  \resizebox{9cm}{!}{\includegraphics{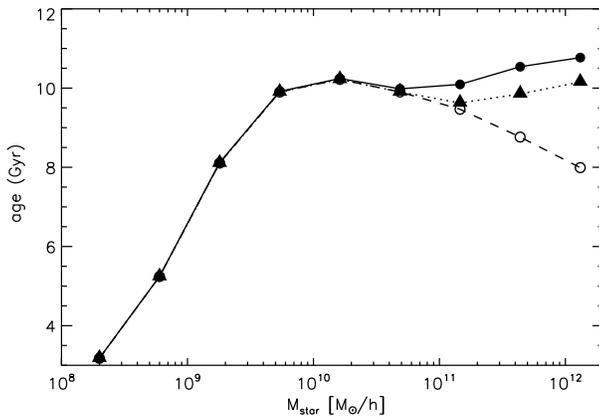}}\\%
\caption{Median age of model elliptical galaxies as a function of galaxy
  stellar mass. Filled circles show results for our default model.  Empty
  circles are for a model without AGN feedback and without any cooling
  cutoff. Filled triangles are for a model without AGN feedback but with a
  cooling cutoff with a critical velocity equal to $350\,{\rm km}\,{\rm
  s}^{-1}$.} 
\label{fig:agecheck}
\ec
\end{figure}

This is better seen in Fig.~\ref{fig:agecheck} where we show the median
luminosity--weighted ages of model elliptical galaxies as a function of their
stellar mass for the same three models.  Filled circles show the result for the
model with AGN feedback, open circles show the result for the model without AGN
feedback and without any artificial cutoff, and filled triangles show the
result for the model with the artificial cooling cutoff.  The lines coincide
perfectly up to stellar masses $\simeq 5\times10^{10}\,{\rm M}_{\sun}$.  For
larger masses, the median age stays almost constant for the model with AGN
feedback, decreases for the model without suppression of the cooling flows, and
shows a non--monotonic behaviour for the model with an artificial cooling
cutoff.  We note, however, that the differences between this scheme and a model
with AGN feedback are small.  In our analysis, a cooling flow cut-off is
hence able to approximately produce the same result as AGN feedback, which is
still different from many earlier results.

We note that the earlier semi-analytic results were based on Monte-Carlo merger
trees constructed with the extended Press-Schechter formalism, and not by
measuring merger trees directly from high-resolution numerical simulations as
we have done here.  Although there have been suggestions that the extended
Press--Schechter theory may not provide a sufficiently accurate description of
the merger trees \citep*{benson05}, it seems unlikely that this is responsible
for reversing the trends found in this analysis.  We believe this reversal to
be a combination of the change in the physical model and in the cosmology.  We
note that the work of \citet{kauffmann98} was carried out in the framework of a
cosmological model with critical matter density, where massive haloes are
formed much later than in a low-density Universe normalised to the same cluster
abundance.  In addition, \citet{kauffmann98} employed an artificial cooling
cutoff corresponding to a critical velocity equal to $500\,{\rm km}\,{\rm
  s}^{-1}$.  This leaves room for late gaseous mergers and gas accretion that
can substantially rejuvenate the stellar population of elliptical galaxies.

Another important difference between our model and previous ones is that we
explicitly follow dark matter substructures within each halo, even after their
progenitor halos have been accreted by larger structures.  \citet{springel01}
have shown that this allows a more faithful tracking of the orbits of galaxies
and an improved modelling of the actual merging rate in any given halo.  Most
of the galaxies we classify as ellipticals are indeed genuine `satellite'
galaxies with their own self-bound dark matter subhalo.  If too many of these
satellite galaxies are assumed to merge on too short a timescale, excessively
bright and blue central galaxies result.

The short formation time--scales we find for very massive ellipticals are
qualitatively in agreement with those required by \citet{thomas04} in order to
reproduce the observed $\alpha$-element enhancements.  The detailed census of
ages and metallicities for stars in our model elliptical galaxies depends on
details of our feedback model and chemical enrichment scheme.  We plan to come
back to these issues in future work. 

Our results demonstrate that an apparent `down-sizing' in the formation of
ellipticals is not in contradiction with the hierarchical paradigm.  Modern
semi-analytic models of galaxy formation do predict `anti-hierarchical' star
formation histories for ellipticals in a $\Lambda$CDM universe even though the
{\it assembly} of these galaxies is indeed hierarchical.
\section*{Acknowledgements}
G.~D.~L. would like to thank M. Pannella and V. Strazzullo for intense and
provocative discussions on our poor knowledge of galaxy formation and
evolution.  We thank B.~Poggianti and A.~Arag{\' o}n-Salamanca for useful
comments and stimulating discussions.  G.~D.~L. thanks the Alexander von
Humboldt Foundation, the Federal Ministry of Education and Research, and the
Programme for Investment in the Future (ZIP) of the German Government for
financial support.

\bsp
\label{lastpage}

\bibliographystyle{mn2e}
\bibliography{delucia_ellip}

\begin{thebibliography}{}

\bibitem[\protect\citeauthoryear{{Barger}, {Aragon-Salamanca}, {Ellis},
  {Couch}, {Smail} \& {Sharples}}{{Barger} et~al.}{1996}]{barger96}
{Barger} A.~J.,  {Aragon-Salamanca} A.,  {Ellis} R.~S.,  {Couch} W.~J.,
  {Smail} I.,    {Sharples} R.~M.,  1996, \mnras, 279, 1

\bibitem[\protect\citeauthoryear{{Baugh}, {Cole} \& {Frenk}}{{Baugh}
  et~al.}{1996}]{baugh96}
{Baugh} C.~M.,  {Cole} S.,    {Frenk} C.~S.,  1996, \mnras, 283, 1361

\bibitem[\protect\citeauthoryear{{Bell}, {McIntosh}, {Katz} \&
  {Weinberg}}{{Bell} et~al.}{2003}]{bell03}
{Bell} E.~F.,  {McIntosh} D.~H.,  {Katz} N.,    {Weinberg} M.~D.,  2003, \apjs,
  149, 289

\bibitem[\protect\citeauthoryear{{Bell}, {Naab}, {McIntosh}, {Somerville},
  {Caldwell}, {Barden}, {Wolf}, {Rix}, {Beckwith}, {Borch}, {Haeussler},
  {Heymans}, {Jahnke}, {Jogee}, {Meisenheimer}, {Peng}, {Sanchez} \&
  {Wisotzki}}{{Bell} et~al.}{2005}]{bell05}
{Bell} E.~F.,  {Naab} T.,  {McIntosh} D.~H.,  {Somerville} R.~S.,  {Caldwell}
  J.~A.~R.,  {Barden} M.,  {Wolf} C.,  {Rix} H.-W.,  {Beckwith} S.~V.~W.,
  {Borch} A.,  {Haeussler} B.,  {Heymans} C.,  {Jahnke} K.,  {Jogee} S.,
  {Meisenheimer} K.,  {Peng} C.~Y.,  {Sanchez} S.~F.,    {Wisotzki} L.,  2005,
  \apj submitted, astro-ph/0506425

\bibitem[\protect\citeauthoryear{{Benson}, {Kamionkowski} \&
  {Hassani}}{{Benson} et~al.}{2005}]{benson05}
{Benson} A.~J.,  {Kamionkowski} M.,    {Hassani} S.~H.,  2005, \mnras, 357, 847

\bibitem[\protect\citeauthoryear{{Bruzual} \& {Charlot}}{{Bruzual} \&
  {Charlot}}{2003}]{BC2003}
{Bruzual} G.,  {Charlot} S.,  2003, \mnras, 344, 1000

\bibitem[\protect\citeauthoryear{{Bruzual A.}}{{Bruzual
  A.}}{1983}]{bruzual1983}
{Bruzual A.} G.,  1983, \apj, 273, 105

\bibitem[\protect\citeauthoryear{{Buzzoni}}{{Buzzoni}}{1989}]{buzzoni1989}
{Buzzoni} A.,  1989, \apjs, 71, 817

\bibitem[\protect\citeauthoryear{{Colless}, {Dalton}, {Maddox}, {Sutherland},
  {Norberg}, {Cole}, {Bland-Hawthorn}, {Bridges} \& {et al.,}}{{Colless}
  et~al.}{2001}]{colless01}
{Colless} M.,  {Dalton} G.,  {Maddox} S.,  {Sutherland} W.,  {Norberg} P.,
  {Cole} S.,  {Bland-Hawthorn} J.,  {Bridges} T.,    {et al.,} 2001, \mnras,
  328, 1039

\bibitem[\protect\citeauthoryear{{Cowie}, {Songaila}, {Hu} \& {Cohen}}{{Cowie}
  et~al.}{1996}]{cowie96}
{Cowie} L.~L.,  {Songaila} A.,  {Hu} E.~M.,    {Cohen} J.~G.,  1996, \aj, 112,
  839

\bibitem[\protect\citeauthoryear{{Croton}, {Springel}, {White}, {De Lucia},
  {Frenk}, {Gao}, {Jenkins}, {Kauffmann}, {Navarro} \& {Yoshida}}{{Croton}
  et~al.}{2005}]{croton05}
{Croton} D.~J.,  {Springel} V.,  {White} S.~D.~M.,  {De Lucia} G.,  {Frenk}
  C.~S.,  {Gao} L.,  {Jenkins} A.,  {Kauffmann} G.,  {Navarro} J.~F.,
  {Yoshida} N.,  2005, \mnras, in press, astro-ph/0508046

\bibitem[\protect\citeauthoryear{{De Lucia}}{{De Lucia}}{2004}]{deluciaPhD}
{De Lucia} G.,  2004, Ph.D.~Thesis, Ludwig-Maximilian Universit\"at, Munich

\bibitem[\protect\citeauthoryear{{De Lucia}, {Kauffmann}, {Springel}, {White},
  {Lanzoni}, {Stoehr}, {Tormen} \& {Yoshida}}{{De Lucia}
  et~al.}{2004a}]{delucia04a}
{De Lucia} G.,  {Kauffmann} G.,  {Springel} V.,  {White} S.~D.~M.,  {Lanzoni}
  B.,  {Stoehr} F.,  {Tormen} G.,    {Yoshida} N.,  2004, \mnras, 348, 333

\bibitem[\protect\citeauthoryear{{De Lucia}, {Kauffmann} \& {White}}{{De Lucia}
  et~al.}{2004b}]{delucia04b}
{De Lucia} G.,  {Kauffmann} G.,    {White} S.~D.~M.,  2004, \mnras, 349, 1101

\bibitem[\protect\citeauthoryear{{De Lucia}, {Poggianti}, {Arag{\'
  o}n-Salamanca}, {Clowe}, {Halliday}, {Jablonka}, {Milvang-Jensen}, {Pell{\'
  o}}, {Poirier}, {Rudnick}, {Saglia}, {Simard} \& {White}}{{De Lucia}
  et~al.}{2004c}]{delucia04c}
{De Lucia} G.,  {Poggianti} B.~M.,  {Arag{\' o}n-Salamanca} A.,  {Clowe} D.,
  {Halliday} C.,  {Jablonka} P.,  {Milvang-Jensen} B.,  {Pell{\' o}} R.,
  {Poirier} S.,  {Rudnick} G.,  {Saglia} R.,  {Simard} L.,    {White} S.~D.~M.,
   2004, \apjl, 610, L77

\bibitem[\protect\citeauthoryear{{Denicol{\'o}}, {Terlevich}, {Terlevich},
  {Forbes} \& {Terlevich}}{{Denicol{\'o}} et~al.}{2005}]{denicolo04}
{Denicol{\'o}} G.,  {Terlevich} R.,  {Terlevich} E.,  {Forbes} D.~A.,
  {Terlevich} A.,  2005, \mnras, 358, 813

\bibitem[\protect\citeauthoryear{{Diaferio}, {Kauffmann}, {Balogh}, {White},
  {Schade} \& {Ellingson}}{{Diaferio} et~al.}{2001}]{diaferio2001}
{Diaferio} A.,  {Kauffmann} G.,  {Balogh} M.~L.,  {White} S.~D.~M.,  {Schade}
  D.,    {Ellingson} E.,  2001, \mnras, 323, 999

\bibitem[\protect\citeauthoryear{{Faber}, {Trager}, {Gonz\'alez} \&
  {Worthey}}{{Faber} et~al.}{1995}]{faber1995}
{Faber} S.~M.,  {Trager} S.,  {Gonz\'alez} J.,    {Worthey} G.,  1995, in van
  der Kruit P.~C., Gilmore G., eds, Stellar Populations. Kluwer, Drdrecht,
  p.~249

\bibitem[\protect\citeauthoryear{{Faber}, {Willmer}, {Wolf}, {Koo}, {Weiner},
  {Newman}, {Im}, {Coil} \& {et}}{{Faber} et~al.}{2005}]{faber05}
{Faber} S.~M.,  {Willmer} C.~N.~A.,  {Wolf} C.,  {Koo} D.~C.,  {Weiner} B.~J.,
  {Newman} J.~A.,  {Im} M.,  {Coil} A.~L.,    {et} a.,  2005, \apj, submitted,
  astro-ph/0506044

\bibitem[\protect\citeauthoryear{{Farouki} \& {Shapiro}}{{Farouki} \&
  {Shapiro}}{1982}]{FS82}
{Farouki} R.~T.,  {Shapiro} S.~L.,  1982, \apj, 259, 103

\bibitem[\protect\citeauthoryear{{Gao}, {Springel} \& {White}}{{Gao}
  et~al.}{2005}]{Gao2005}
{Gao} L.,  {Springel} V.,    {White} S.~D.~M.,  2005, \mnras, 363, L66

\bibitem[\protect\citeauthoryear{{Gao}, {White}, {Jenkins}, {Stoehr} \&
  {Springel}}{{Gao} et~al.}{2004}]{gao04a}
{Gao} L.,  {White} S.~D.~M.,  {Jenkins} A.,  {Stoehr} F.,    {Springel} V.,
  2004, \mnras, 355, 819

\bibitem[\protect\citeauthoryear{{Guiderdoni} \&
  {Rocca-Volmerange}}{{Guiderdoni} \&
  {Rocca-Volmerange}}{1987}]{guiderdoni1987}
{Guiderdoni} B.,  {Rocca-Volmerange} B.,  1987, \aap, 186, 1

\bibitem[\protect\citeauthoryear{{Kauffmann}}{{Kauffmann}}{1995}]{kauffmann95}
{Kauffmann} G.,  1995, \mnras, 274, 161

\bibitem[\protect\citeauthoryear{{Kauffmann}}{{Kauffmann}}{1996a}]{kauffmann96%
b}
{Kauffmann} G.,  1996a, \mnras, 281, 475

\bibitem[\protect\citeauthoryear{{Kauffmann}}{{Kauffmann}}{1996b}]{kauffmann96%
a}
{Kauffmann} G.,  1996b, \mnras, 281, 487

\bibitem[\protect\citeauthoryear{{Kauffmann} \& {Charlot}}{{Kauffmann} \&
  {Charlot}}{1998}]{kauffmann98}
{Kauffmann} G.,  {Charlot} S.,  1998, \mnras, 294, 705

\bibitem[\protect\citeauthoryear{{Kauffmann}, {Colberg}, {Diaferio} \&
  {White}}{{Kauffmann} et~al.}{1999}]{kauffmann99}
{Kauffmann} G.,  {Colberg} J.~M.,  {Diaferio} A.,    {White} S.~D.~M.,  1999,
  \mnras, 303, 188

\bibitem[\protect\citeauthoryear{{Kodama}, {Arimoto}, {Barger} \&
  {Arag\'on-Salamanca}}{{Kodama} et~al.}{1998}]{kodama98}
{Kodama} T.,  {Arimoto} N.,  {Barger} A.~J.,    {Arag\'on-Salamanca} A.,  1998,
  \aap, 334, 99

\bibitem[\protect\citeauthoryear{{Kodama}, {Yamada}, {Akiyama}, {Aoki}, {Doi},
  {Furusawa}, {Fuse}, {Imanishi} \& {et}}{{Kodama} et~al.}{2004}]{kodama2004}
{Kodama} T.,  {Yamada} T.,  {Akiyama} M.,  {Aoki} K.,  {Doi} M.,  {Furusawa}
  H.,  {Fuse} T.,  {Imanishi} M.,    {et} a.,  2004, \mnras, 350, 1005

\bibitem[\protect\citeauthoryear{{Larson}}{{Larson}}{1975}]{larson75}
{Larson} R.~B.,  1975, \mnras, 173, 671

\bibitem[\protect\citeauthoryear{{Loveday}}{{Loveday}}{1996}]{loveday96}
{Loveday} J.,  1996, \mnras, 278, 1025

\bibitem[\protect\citeauthoryear{{Mathis}, {Lemson}, {Springel}, {Kauffmann},
  {White}, {Eldar} \& {Dekel}}{{Mathis} et~al.}{2002}]{mathis02}
{Mathis} H.,  {Lemson} G.,  {Springel} V.,  {Kauffmann} G.,  {White} S.~D.~M.,
  {Eldar} A.,    {Dekel} A.,  2002, \mnras, 333, 739

\bibitem[\protect\citeauthoryear{{Menanteau}, {Abraham} \& {Ellis}}{{Menanteau}
  et~al.}{2001}]{menanteau01}
{Menanteau} F.,  {Abraham} R.~G.,    {Ellis} R.~S.,  2001, \mnras, 322, 1

\bibitem[\protect\citeauthoryear{{Michard} \& {Prugniel}}{{Michard} \&
  {Prugniel}}{2004}]{michard04}
{Michard} R.,  {Prugniel} P.,  2004, \aap, 423, 833

\bibitem[\protect\citeauthoryear{{Mo}, {Mao} \& {White}}{{Mo}
  et~al.}{1998}]{mo1998}
{Mo} H.~J.,  {Mao} S.,    {White} S.~D.~M.,  1998, \mnras, 295, 319

\bibitem[\protect\citeauthoryear{{Negroponte} \& {White}}{{Negroponte} \&
  {White}}{1983}]{NW83}
{Negroponte} J.,  {White} S.~D.~M.,  1983, \mnras, 205, 1009

\bibitem[\protect\citeauthoryear{{Nelan}, {Smith}, {Hudson}, {Wegner}, {Lucey},
  {Moore}, {Quinney} \& {Suntzeff}}{{Nelan} et~al.}{2005}]{nelan05}
{Nelan} J.~E.,  {Smith} R.~J.,  {Hudson} M.~J.,  {Wegner} G.~A.,  {Lucey}
  J.~R.,  {Moore} S.~A.~W.,  {Quinney} S.~J.,    {Suntzeff} N.~B.,  2005, \apj,
  632, 137

\bibitem[\protect\citeauthoryear{{Partridge} \& {Peebles}}{{Partridge} \&
  {Peebles}}{1967}]{PP67}
{Partridge} R.~B.,  {Peebles} P.~J.~E.,  1967, \apj, 147, 868

\bibitem[\protect\citeauthoryear{{Proctor}, {Forbes}, {Hau}, {Beasley}, {De
  Silva}, {Contreras} \& {Terlevich}}{{Proctor} et~al.}{2004}]{proctor04}
{Proctor} R.~N.,  {Forbes} D.~A.,  {Hau} G.~K.~T.,  {Beasley} M.~A.,  {De
  Silva} G.~M.,  {Contreras} R.,    {Terlevich} A.~I.,  2004, \mnras, 349, 1381

\bibitem[\protect\citeauthoryear{{Schweizer} \& {Seitzer}}{{Schweizer} \&
  {Seitzer}}{1992}]{schweizer92}
{Schweizer} F.,  {Seitzer} P.,  1992, \aj, 104, 1039

\bibitem[\protect\citeauthoryear{{Simien} \& {de Vaucouleurs}}{{Simien} \& {de
  Vaucouleurs}}{1986}]{simien}
{Simien} F.,  {de Vaucouleurs} G.,  1986, \apj, 302, 564

\bibitem[\protect\citeauthoryear{{Somerville}, {Primack} \&
  {Faber}}{{Somerville} et~al.}{2001}]{som01}
{Somerville} R.~S.,  {Primack} J.~R.,    {Faber} S.~M.,  2001, \mnras, 320, 504

\bibitem[\protect\citeauthoryear{{Spergel}, {Verde}, {Peiris}, {Komatsu},
  {Nolta}, {Bennett}, {Halpern}, {Hinshaw}, {Jarosik}, {Kogut}, {Limon},
  {Meyer}, {Page}, {Tucker}, {Weiland}, {Wollack} \& {Wright}}{{Spergel}
  et~al.}{2003}]{spergel03}
{Spergel} D.~N.,  {Verde} L.,  {Peiris} H.~V.,  {Komatsu} E.,  {Nolta} M.~R.,
  {Bennett} C.~L.,  {Halpern} M.,  {Hinshaw} G.,  {Jarosik} N.,  {Kogut} A.,
  {Limon} M.,  {Meyer} S.~S.,  {Page} L.,  {Tucker} G.~S.,  {Weiland} J.~L.,
  {Wollack} E.,    {Wright} E.~L.,  2003, \apjs, 148, 175

\bibitem[\protect\citeauthoryear{{Springel}, {White}, {Jenkins}, {Frenk},
  {Yoshida}, {Gao}, {Navarro}, {Thacker}, {Croton}, {Helly}, {Peacock}, {Cole},
  {Thomas}, {Couchman}, {Evrard}, {Colberg} \& {Pearce}}{{Springel}
  et~al.}{2005}]{springel2005}
{Springel} V.,  {White} S.~D.~M.,  {Jenkins} A.,  {Frenk} C.~S.,  {Yoshida} N.,
   {Gao} L.,  {Navarro} J.,  {Thacker} R.,  {Croton} D.,  {Helly} J.,
  {Peacock} J.~A.,  {Cole} S.,  {Thomas} P.,  {Couchman} H.,  {Evrard} A.,
  {Colberg} J.,    {Pearce} F.,  2005, Nature, 435, 629

\bibitem[\protect\citeauthoryear{{Springel}, {White}, {Tormen} \&
  {Kauffmann}}{{Springel} et~al.}{2001}]{springel01}
{Springel} V.,  {White} S.~D.~M.,  {Tormen} G.,    {Kauffmann} G.,  2001,
  \mnras, 328, 726

\bibitem[\protect\citeauthoryear{{Terlevich} \& {Forbes}}{{Terlevich} \&
  {Forbes}}{2002}]{terlevich02}
{Terlevich} A.~I.,  {Forbes} D.~A.,  2002, \mnras, 330, 547

\bibitem[\protect\citeauthoryear{{Thomas}}{{Thomas}}{1999}]{thomas99}
{Thomas} D.,  1999, \mnras, 306, 655

\bibitem[\protect\citeauthoryear{{Thomas}, {Maraston} \& {Bender}}{{Thomas}
  et~al.}{2003}]{TMB2003}
{Thomas} D.,  {Maraston} C.,    {Bender} R.,  2003, \mnras, 339, 897

\bibitem[\protect\citeauthoryear{{Thomas}, {Maraston}, {Bender} \& {de
  Oliveira}}{{Thomas} et~al.}{2005}]{thomas04}
{Thomas} D.,  {Maraston} C.,  {Bender} R.,    {de Oliveira} C.~M.,  2005, \apj,
  621, 673

\bibitem[\protect\citeauthoryear{{Tinsley}}{{Tinsley}}{1972}]{tinsley1972}
{Tinsley} B.~M.,  1972, \aap, 20, 383

\bibitem[\protect\citeauthoryear{{Toomre} \& {Toomre}}{{Toomre} \&
  {Toomre}}{1972}]{toomre72}
{Toomre} A.,  {Toomre} J.,  1972, \apj, 178, 623

\bibitem[\protect\citeauthoryear{{Tran}, {van Dokkum}, {Franx}, {Illingworth},
  {Kelson} \& {Schreiber}}{{Tran} et~al.}{2005}]{tran05}
{Tran} K.-V.~H.,  {van Dokkum} P.,  {Franx} M.,  {Illingworth} G.~D.,  {Kelson}
  D.~D.,    {Schreiber} N.~M.~F.,  2005, \apjl, 627, L25

\bibitem[\protect\citeauthoryear{{Treu}, {Ellis}, {Liao} \& {van
  Dokkum}}{{Treu} et~al.}{2005}]{treu05}
{Treu} T.,  {Ellis} R.~S.,  {Liao} T.~X.,    {van Dokkum} P.~G.,  2005, \apjl,
  622, L5

\bibitem[\protect\citeauthoryear{{Treu}, {Stiavelli}, {Casertano}, {M{\o}ller}
  \& {Bertin}}{{Treu} et~al.}{2002}]{treu02}
{Treu} T.,  {Stiavelli} M.,  {Casertano} S.,  {M{\o}ller} P.,    {Bertin} G.,
  2002, \apjl, 564, L13

\bibitem[\protect\citeauthoryear{{van de Ven}, {van Dokkum} \& {Franx}}{{van de
  Ven} et~al.}{2003}]{vandeven03}
{van de Ven} G.,  {van Dokkum} P.~G.,    {Franx} M.,  2003, \mnras, 344, 924

\bibitem[\protect\citeauthoryear{{van der Wel}, {Franx}, {van Dokkum}, {Rix},
  {Illingworth} \& {Rosati}}{{van der Wel} et~al.}{2005}]{vandervel05}
{van der Wel} A.,  {Franx} M.,  {van Dokkum} P.~G.,  {Rix} H.-W.,
  {Illingworth} G.~D.,    {Rosati} P.,  2005, \apj, 631, 145

\bibitem[\protect\citeauthoryear{{van Dokkum}}{{van Dokkum}}{2005}]{dokkum05}
{van Dokkum} P.~G.,  2005, \aj, in press, astro-ph/0506661

\bibitem[\protect\citeauthoryear{{van Dokkum} \& {Franx}}{{van Dokkum} \&
  {Franx}}{1996}]{dokkumfranx1996}
{van Dokkum} P.~G.,  {Franx} M.,  1996, \mnras, 281, 985

\bibitem[\protect\citeauthoryear{{van Dokkum} \& {Stanford}}{{van Dokkum} \&
  {Stanford}}{2003}]{vandokkum03}
{van Dokkum} P.~G.,  {Stanford} S.~A.,  2003, \apj, 585, 78

\bibitem[\protect\citeauthoryear{{Vazdekis}}{{Vazdekis}}{2001}]{vazdekis2001}
{Vazdekis} A.,  2001, \apss, 276, 921

\bibitem[\protect\citeauthoryear{{Willis}, {Hewett}, {Warren} \&
  {Lewis}}{{Willis} et~al.}{2002}]{willis02}
{Willis} J.~P.,  {Hewett} P.~C.,  {Warren} S.~J.,    {Lewis} G.~F.,  2002,
  \mnras, 337, 953

\end{thebibliography}

\end{document}